\documentstyle[aps,twocolumn,epsfig]{revtex}

\begin{document}
\title{Nonclassical photon pairs generated from a room-temperature atomic ensemble}
\author{Wei Jiang$^{1}$, Chao Han$^{1}$, Peng Xue$^{1}$, L. -M. Duan$^{1,2}$, G. -C.
Guo$^{1}$}
\address{$^{1}$Key Laboratory of Quantum Information, University of Science\\
and\\
Technology of China, Hefei 230026, P. R. China\\
$^{2}$Department of Physics, University of Michigan, Ann Arbor, MI 48109-1120}
\maketitle

\begin{abstract}
We report experimental generation of non-classically correlated photon pairs
from collective emission in a room-temperature atomic vapor cell. The
nonclassical feature of the emission is demonstrated by observing a
violation of the Cauchy-Schwarz inequality. Each pair of correlated photons
are separated by a controllable time delay up to $2$ microseconds. This
experiment demonstrates an important step towards the realization of the
Duan-Lukin-Cirac-Zoller scheme for scalable long-distance quantum
communication.

\end{abstract}

Implementation of quantum communication and computation is of large current
interest \cite{0}. Recently, significant interests arise in using atomic
ensembles for realization of spin squeezing \cite{1,2,3,4}, quantum memory 
\cite{5,6,7,8}, and quantum information processing \cite{9,10,11,12,Duan,14}%
. A significant result in this direction is a scheme proposed by Duan,
Lukin, Cirac and Zoller (hereafter ``DLCZ'') which gives a promising
approach to realization of scalable long-distance quantum communication \cite
{Duan}. This scheme overcomes the photon attenuation problem and can
communicate quantum states over arbitrarily long distance with only
polynomial costs based on a clever implementation of the quantum repeater
architecture \cite{15,16}. In the DLCZ scheme, one generates and purifies
entanglement between distant atomic ensembles by interfering and counting
single-photons emitted from them. The same setup can also be used to prepare
maximally entanglement states between many such ensembles \cite{14}. The
challenging and important enabling step for these schemes is to demonstrate
quantum correlation between the emitted single-photons and the long-lived
collective atomic excitations. In this experiment, we realize this important
step by observing non-classical correlation between them in a
room-temperature atomic vapor. The collective atomic excitation is
transferred subsequently to a photon in experiments, so actually we observe
nonclassical correlation between two successively emitted photons.

Very recently, two other experiments have been reported contributing to the
realization of the DLCZ scheme \cite{Kuzmich,18}. In particular, Kuzmich et
al. have reported striking results on generation of nonclassical photon
pairs in a sample of cold atoms. Compared with those experiments, our
experiment is distinctive by the following features: (i) We observe
nonclassical correlation between photon pairs generated from a
room-temperature atomic vapor cell. The observed correlation is comparable
to that reported in Ref. \cite{Kuzmich}, but our setup of room-temperature
atoms is much cheaper compared with the magnetic-optical trap used for cold
atoms \cite{Kuzmich}, and the use of cheaper setups could have advantage in
future full realization of the DLCZ\ scheme where many such basic setups are
required. The experiment in Ref. \cite{18} is also based on the use of an
atomic vapor cell, however, it is not performed in the single-photon region
as required by the DLCZ scheme. (ii) Compared with instantaneous photon
pairs produced previously in atomic cascades \cite{Clauser} or in parametric
down conversion \cite{20}, the atomic ensemble experiments are distinct in
that the correlated photon pairs generated in this system can be separated
with a controllable time delay, which is in principle only limited by the
single-photon storage time in the atomic ensemble. In our experiment, we
demonstrate a time delay of $2$ microseconds between the pair of
non-classsically correlated photons, which is considerably longer than the $%
400$-nanosecond time delay reported before \cite{Kuzmich}. It is important
to experimentally improve this storage time to facilitate the full
realization of the DLCZ scheme which requires quantum memory \cite{Duan,14}.

The basic idea behind this experiment can be understood by considering a
sample of three-level atoms in a $\Lambda $ type configuration as
illustrated in Fig. 1b. Most of the atoms are initially prepared in the
state $\left| a\right\rangle $ through optical pumping. A write laser pulse
is then sent through the atomic ensemble which couples off-resonantly to the
atomic transition $\left| a\right\rangle \rightarrow \left| e\right\rangle $%
. This pulse induces a Raman scattering, bringing a small fraction of the
atoms into the level $\left| b\right\rangle $ by emitting Stokes photons
from the transition $\left| e\right\rangle \rightarrow \left| b\right\rangle 
$. The write pulse is controlled to be weak so that the average Stokes
photon number scattered into the specified forward propagating mode $\psi
_{w}\left( {\bf r}\right) $ is much smaller than $1$ for each pulse \cite
{Duan,21}. We detect the photon in this mode, and upon a detector click, one
atom will be excited to the collective atomic mode $s^{\dagger }$ defined as 
\cite{Duan,21} 
\begin{equation}
s^{\dagger }=\left( 1/\sqrt{N_{a}}\right) \sum_{i=1}^{N_{a}}\left|
b\right\rangle _{i}\left\langle a\right| ,
\end{equation}
where $N_{a}$ is the total involved atom number. It has been predicted that
there will be a definite correlation between the photon number in the
forward propagating mode $\psi _{1}\left( {\bf r}\right) $ and the atom
number in the collective mode $s$ \cite{Duan}, and this correlation is
critical for all the applications based on this setup, including the
implementation of quantum repeaters. To experimentally confirm this
correlation, we transfer the collective atomic excitation to a photonic
excitation after a time delay $\delta t$ by shining a read laser pulse on
the ensemble which couples to the transition $\left| b\right\rangle
\rightarrow \left| e\right\rangle $. This read pulse will bring the atom
back to the state $\left| a\right\rangle $ by emitting an anti-Stokes photon
in a specified forward-propagating mode $\psi _{r}\left( {\bf r}\right) $.
We can then detect the photon coincidences in the write and the read modes $%
\psi _{w}\left( {\bf r}\right) $ and $\psi _{r}\left( {\bf r}\right) $. The
mode function $\psi _{r}\left( {\bf r}\right) $ is determined by the spatial
shapes of the write and the read pulses, the geometry of the atomic
ensemble, and the mode structure of $\psi _{w}\left( {\bf r}\right) $ \cite
{21}. In experiments, one can choose the waists of the write and the read
laser beams so that $\psi _{w}\left( {\bf r}\right) $ and $\psi _{r}\left( 
{\bf r}\right) $ largely overlap with the spatial shapes of these pumping
beams. The Stokes and anti-Stokes photons in these modes are coupled into
single-mode fibers, which direct them to single-photon detectors for
coincidence measurements.

In our experiment, we use an optically thick ensemble of $^{87}$Rb atoms.
The basic element is an isotopically pure $^{87}$Rb atomic vapor contained
in a silica cell. There is also some buffer gas (Ne, $30$ Torr) inside the
cell which is used to increase the spin relaxation time of the $^{87}$Rb
atoms. The number density of $^{87}$Rb atoms is estimated to be around $%
1.3\times 10^{10}$ cm$^{-3}$ under room temperature. The silica cell is
placed in a three-layer permalloy box for shielding magnetic fields. The
residue magnetic field inside the box is estimated to be below $1$ mG. The
desired ground states $\left| a\right\rangle $ and $\left| b\right\rangle $
are chosen respectively as the hyperfine states $\left|
5S_{1/2},F=1\right\rangle $ and $\left| 5S_{1/2},F=2\right\rangle $ of the $%
^{87}$Rb atoms, and the excited state $\left| e\right\rangle $ is provided
by the hyperfine states in the $\left| 5P_{1/2}\right\rangle $ manifold.

The schematic set up for this experiment is shown by Fig. 1a. The write and
the read pulses are from two different semiconductor lasers working at a
wavelength about $795$ nm, with the frequency difference of $6.8$ GHz
matching the hyperfine splitting of the $^{87}$Rb atoms. These pumping laser
beams are shined from different sides of the first polarization beam
splitter PBS$_{1}$ so that they have orthogonal polarizations when going
through the silica atomic vapor cell. After the cell, we need to separate
the weak signal of Stokes or anti-Stokes photons from the strong write and
read laser pulses, and this is done through both polarization and frequency
selection. Right after the cell, the second polarized beam splitter PBS$_{2}$
will separate the signal from the pumping laser beams with an extinction
ratio of about $7\times 10^{3}$. Further frequency filtering is achieved by
the glass cells F1 and F2, each containing Rubidium atoms initially
optically pumped to the hyperfine levels $\left| 5S_{1/2},F=1\right\rangle $
and $\left| 5S_{1/2},F=2\right\rangle $, respectively. As in Ref. \cite
{Kuzmich}, the residual write (read) laser pulses after the PBS$_{2}$ will
be strongly attenuated $\left( >10^{5}\right) $ by the atomic cells F1 (F2)
through resonant absorption while the signal Stokes (anti-Stokes) photons
transmit with a high efficiency due to the large hyperfine detuning. After
the filters, both of the Stokes and anti-Stokes photons are split by a $%
50\%-50\%$ beam splitter, and then coupled into single-mode fibers, which
direct them to the four single photon detectors (PerkinElmer Model
SPCM-AQR). With this setup, we can measure both the auto-correlations and
cross-correlations between the Stokes and the anti-Stokes photons.

The experiment goes as follows: we first optically pump the $^{87}$Rb atoms
in the silica cell to the ground state $\left| 5S_{1/2},F=1\right\rangle $.
Then the write pulse, with about $5\times 10^3$ photons, is sent through the
atomic ensemble. The duration of the write pulse is about $1\mu s$. With
this pulse, the probability to generate a Stokes photon in the spatial mode
collected in this experiment is estimated to be about $p_w\approx 0.14$. The
Stokes photons are detected by the single photon detectors $D_A$ and $D_B$.
After a controllable time delay $\delta t$, which is typically $2$ $\mu s$
for our experiment, we send the read pulse through the atomic ensemble. This
pulse has the same duration as the write pulse, but is stronger in intensity
than the latter by a factor of $10$. This read pulse transfers the
collective atomic excitations back to anti-Stokes photons, with the
retrieving probability of about $0.32$ estimated from our experimental data
(including the half polarization loss due to the unpolarized ensemble). The
anti-Stokes photons are registered through the single-photon detectors $D_C$
and $D_D$. To reduce noise, before the detectors $\left( D_A,D_B\right) $
and $\left( D_C,D_D\right) $, we apply two gates with the time window of
about $1$ $\mu s$, synchronized respectively with the write and the read
pulses. The time sequences of the write and the read pulses together with
the gating windows are shown by Fig. 1c. The above steps form a full duty
cycle (one trial), and this cycle is repeated at a rate of $5$ kHz.

In experiments, the detectors for the Stokes and anti-Stokes photons
typically have count rates of $220$ s$^{-1}$ and $70$ s$^{-1}$,
respectively. The outputs of these four single photon detectors are sent to
a time interval analyzer (TIA) for measuring the coincidence between any
chosen pair of detectors. This coincidence can be measured by\ using the
output from one detector as the start signal of the TIA and recording the
arrival time of the output from the other detector$.$ The coincidence
between $\left( D_{A},D_{B}\right) $, $\left( D_{C},D_{D}\right) $, and $%
\left( D_{A},D_{C}\right) $ (or $\left( D_{B},D_{D}\right) $) are denoted in
the following by $n_{1,1}(t),$ $n_{2,2}(t),$ and $n_{1,2}(t)$, respectively.
The experimental results for these coincidences are shown by Fig. 2a (the
left column). The first peak represents the coincidence within the same duty
cycle (say $i$), and the following peaks are coincidences between the $i$th
trial and the following trials. In Fig. 2b (the right column), by expanding
the time axis, we show the detailed time shape $n_{\alpha ,\beta }(\tau )$
of the first peak as well as the average time shape $m_{\alpha ,\beta }(\tau
)$ of the seven following peaks, where $\alpha ,\beta =1,2$, and $\tau $
denotes the shifted arrival time whose width is smaller than $1\mu s$ as set
by the gating window. The average shape $m_{\alpha ,\beta }(\tau )$ of the
coincidence rates from different trials are defined as $m_{\alpha ,\beta
}(\tau )=\left( 1/7\right) \sum_{j=1}^{7}$ $n_{\alpha ,\beta }(\tau +j\Delta
t)$, where $\Delta t=200\mu s$ is the time interval between subsequent duty
cycles.

From these measured time resolved coincidences, we can confirm the
nonclassical (quantum) correlation between the Stokes and the anti-Stokes
fields. As the anti-Stokes field is transferred from the collective atomic
mode, this also confirms the nonclassical correlation between the collective
atomic mode and the forward propagating Stokes mode. To confirm the
nonclassical feature of the correlation between the Stokes and anti-Stokes
fields, as in Ref. \cite{Kuzmich} we make use of the Cauchy-Schwarz
inequality. As has been pointed out by Clauser \cite{Clauser} and discussed
in detail in ref. \cite{Kuzmich}, the normalized auto and cross correlations
functions $g_{1,1},$ $g_{2,2},$ and $g_{1,2}$ between two arbitrary
classical fields 1 and 2 need to satisfy the following Cauchy-Schwarz
inequality (here ``classical'' means that there exists P-representation for
these fields with a positive distribution \cite{22}) 
\begin{equation}
\left[ g_{1,2}\right] ^{2}\leq g_{1,1}g_{2,2}.
\end{equation}
The auto and cross correlation functions between the Stokes filed 1 and the
anti-Stokes filed 2 can be directly obtained from the measured coincidence
rates $n_{\alpha ,\beta }(\tau )$\ and $m_{\alpha ,\beta }(\tau )$ $\left(
\alpha ,\beta =1,2\right) $. Let $N_{\alpha ,\beta }$\ and $M_{\alpha ,\beta
}$ represent the total number of coincidences within the time gating window
for the same trial and different trails, respectively, so they actually
represent respectively the time integrals of $n_{\alpha ,\beta }(\tau )$\
and $m_{\alpha ,\beta }(\tau )$ over the whole peak, i.e., they are
determined by the areas of the corresponding peaks. By definition, the
normalized auto and cross correlation functions $g_{1,1},$ $g_{2,2},$ and $%
g_{1,2}$ between the Stokes and anti-Stokes fields are given respectively by 
$g_{1,1}=N_{1,1}/M_{1,1}$, $g_{2,2}=N_{2,2}/M_{2,2}$, and $g_{1,2}\left(
\delta t\right) =N_{1,2}/M_{1,2}$, where $\delta t=2$ $\mu s$ denotes
explicitly the time delay between the Stokes and ani-Stokes fields.

The integrated coincidence rates $N_{\alpha ,\beta }$\ and $M_{\alpha ,\beta
}$ in our experiment are shown by Fig. 2c, from which we calculate the
correlations $g_{1,1}=1.764\pm 0.026,$ $g_{2,2}=1.771\pm 0.028,$ $%
g_{1,2}(\delta t)=2.043\pm 0.031.$ One can see that $\left[ g_{1,2}^2(\delta
t)=4.17\pm 0.09\right] >\left[ g_{1,1}g_{2,2}=3.12\pm 0.08\right] $, so the
Cauchy-Schwarz inequality (2) is manifestly violated by our experiment. This
clearly demonstrates that up to a delay time of $2$ $\mu s$, we still get
non-classically (quantum) correlated photon pairs from our experiment$.$

In the ideal case, if there are no noise and imperfections, the
Cauchy-Schwarz inequality could be violated to a much larger extent. If the
excitation probability of the collective atomic mode is $p$ for each write
pumping pulse, we could get a violation of the inequality (2) with $\left[
g_{1,2}^{2}/g_{1,1}g_{2,2}\right] \simeq \left[ (1+p)/\left( 2p\right)
\right] ^{2}$ in the ideal case. In our experiment, the excitation
probability $p$ is estimated to be between $0.1$ and $0.2$, which in
principle could allow a significantly larger violation. In practice,
however, several sources of noise and imperfection degrade the extent of
violation of the Cauchy-Schwarz inequality. Firstly, in a room-temperature
atomic vapor, due to the atomic motion and the Doppler broadening, the
atomic excitation can be diffused from the collective atomic mode to some
other modes, and vice versa. Such a diffusion will significantly reduce the
cross-correlation between the Stokes and anti-Stokes fields. Secondly, there
are significant background fields due to imperfect filtering. The background
fields are uncorrelated and they will also reduce the cross-correlation
between the Stokes and anti-Stokes fields. Unlike the cold atom ensemble 
\cite{Kuzmich}, the optical thickness of the atomic vapor cell is pretty
large, so we estimate that uncorrelated spontaneous emission is not a
dominant source of noise in our case \cite{21}. The signal-to-noise ratio of
this experiment probably could be improved by increasing the detuning of the
pumping laser, by improving the filtering ratios, by reducing the power of
the write beam, or by adjusting the configuration of the atomic cell to
prolong the spin relaxation time.

In summary, we have observed generation of non-classical photon pairs from a
room-temperature atomic vapor cell. Each pair of correlated photons can be
separated by a controllable time delay which is limit only by the coherence
time in the atomic ground state manifold. Up to a delay time of $2$ $\mu s$,
we still clearly demonstrate non-classical correlation between the Stokes
and anti-Stokes fields from the room-temperature atomic vapor. Compared with
the cold atom experiment, the setup involved here are much cheaper. This
experiment shows the prospect that such a cheap system could be used as a
basic element for realization of quantum repeaters and long-distance quantum
communication.

{\bf Acknowledgment:} We thank Shu-Yu Zhou and Yu-Zhu Wang (Shanghai
Institute of Optics) for their enormous help on this experiment. We also
thank Alex Kuzmich and Jeff Kimble for the helpful discussions. This work
was funded by National Fundamental Research Program ( 2001CB309300),
National Natural Science Foundation of China, the Innovation funds from
Chinese Academy of Sciences, the outstanding Ph. D thesis award and the
CAS's talented scientist award.

\begin{figure}[tbp]
\caption{The experimental configuration. {\bf (1a) }The schematic setup of
the experiment. The write and the read laser beams are sent from different
sides of the PBS$_{1}$, and go through the atomic cell $C$ with orthogonal
polarizations. The silica cell $C$ has a length of $3$ cm, containing
isotopically pure $^{87}$Rb atomic vapor together with Ne buffer gas. The
waist of the laser beams at the cell is about $4$ mm. After the cell, the
signals are separated from the strong pumping laser beams first through the
polarization selection at the PBS$_{2}$ and then through the
frequency-selective absorption at the filter cells $F_{1}$ and $F_{2}$. The
signals are split by a beam splitter, and coupled into single-mode fibers
with an effective collection solid angle of about $2\times 10^{-5}$. Then
they are detected by four single-photon detectors with a detection
efficiency of $64\%$. The transmission efficiency of the signal photons from
the atomic cell to the detectors is estimated to be about $50\%$. The
outputs of the detectors are sent to the time interval analyzer for
measuring the coincidences of any pair of detectors, from which we can infer
both auto-correlation and cross-correlation between the Stokes and
anti-Stokes photons. {\bf (1b)} The relevant level structure of $^{87}$Rb
for this experiment. The desired $\Lambda $ configuration is formed by the
three hyperfine levels $\left| 5S_{1/2},F=1\right\rangle $, $\left|
5S_{1/2},F=2\right\rangle $and $\left| 5P_{1/2},F=1\right\rangle $. The
frequencies of the write and the read laser beams are detuned from the
frequencies of the corresponding atomic transitions both by a detuning of
about $100$ MHz. {\bf (1c)} Time sequences for the optical pump, the write,
and the read pulses, and for the synchronized gating windows. Gate1 and Gate
2 for the write and the read steps both have a time width of about $1$ $\mu
s $.}
\end{figure}

\begin{figure}[tbp]
\caption{The experimental data. ({\bf 2a}) (the left column) The time
resolved coincidence rates $n_{\alpha ,\beta }(t)$ with $(\alpha ,\beta
)=(1,1)$, $(2,2)$ and $(1,2)$ over $8$ successive trials of the experiment. (%
{\bf 2b}) (the right column) The shape of the coincidence peaks with the
time axis expanded. The higher peak corresponds to the coincidence within
the same $i$th trial, and the lower peak represents the average of the
following $7$ coincidence peaks for different trials. ({\bf 2c}) The time
integrated (total) coincidence events from different pairs of detectors.}
\end{figure}

\end{document}